\documentstyle[11pt,aaspp4,rotate]{article}
\begin{document}

\slugcomment{AJ in press}

\title{A Survey of $z>5.7$ Quasars in the Sloan Digital Sky Survey IV:
Discovery of Seven Additional Quasars\altaffilmark{1, 2}}

\author{Xiaohui Fan\altaffilmark{\ref{Arizona},\ref{KPNO}},
Michael A. Strauss\altaffilmark{\ref{Princeton}},
Gordon T. Richards\altaffilmark{\ref{Princeton},\ref{JHU}},
Joseph F. Hennawi\altaffilmark{\ref{Berkeley}},
Robert H. Becker\altaffilmark{\ref{UCDavis},\ref{IGPP}},
Richard L. White\altaffilmark{\ref{STScI}},
Aleksandar M. Diamond-Stanic\altaffilmark{\ref{Arizona}},
Jennifer L. Donley\altaffilmark{\ref{Arizona},\ref{KPNO}},
Linhua Jiang\altaffilmark{\ref{Arizona},\ref{KPNO}},
J. Serena Kim\altaffilmark{\ref{Arizona}},
Marianne Vestergaard\altaffilmark{\ref{Arizona}},
Jason E. Young\altaffilmark{\ref{Arizona},\ref{KPNO}},
James E. Gunn\altaffilmark{\ref{Princeton}},
Robert H. Lupton\altaffilmark{\ref{Princeton}}, 
Gillian R. Knapp\altaffilmark{\ref{Princeton}}, 
Donald P. Schneider\altaffilmark{\ref{PSU}}, 
W. N. Brandt\altaffilmark{\ref{PSU}},
Neta A. Bahcall\altaffilmark{\ref{Princeton}},
J.C. Barentine\altaffilmark{\ref{APO}},
J. Brinkmann\altaffilmark{\ref{APO}},
Howard J. Brewington\altaffilmark{\ref{APO}},
Masataka Fukugita\altaffilmark{\ref{CosmicRay}},
Michael Harvanek\altaffilmark{\ref{APO}},
S.J. Kleinman\altaffilmark{\ref{APO}},
Jurek Krzesinski\altaffilmark{\ref{APO},\ref{Cracow}},
Dan Long\altaffilmark{\ref{APO}},
Eric H. Neilsen, Jr.\altaffilmark{\ref{FNAL}},
Atsuko Nitta\altaffilmark{\ref{APO}},
Stephanie A. Snedden\altaffilmark{\ref{APO}} 
and Wolfgang Voges\altaffilmark{\ref{MPE}}
}

\altaffiltext{1}{Based on observations obtained with the
Sloan Digital Sky Survey,
and with the Apache Point Observatory
3.5-meter telescope,
which is owned and operated by the Astrophysical Research Consortium;
with the MMT Observatory, a joint facility of the University of
Arizona and the Smithsonian Institution; 
with the University of Arizona 2.3-meter Bok Telescope; 
with the Kitt Peak National Observatory 4-meter Mayall Telescope, 
with the 6.5-meter Walter Baade Telescope at the
Las Campanas Observatory, a collaboration
between the Observatories of the Carnegie Institution of Washington, University of Arizona, Harvard University, University of Michigan, and Massachusetts Institute of Technology, 
and at the W.M. Keck Observatory, which is operated as a scientific partnership
among the California Institute of Technology, the University of California and the National Aeronautics and
Space Administration, made possible by the generous financial support of the W.M. Keck 
Foundation.}
\altaffiltext{2}{
This paper is dedicated to the memory of John N. Bahcall.
}

\newcounter{address}
\setcounter{address}{3}
\altaffiltext{\theaddress}{Steward Observatory, The University of Arizona,
Tucson, AZ 85721
\label{Arizona}}
\addtocounter{address}{1}
\altaffiltext{\theaddress}{
Visiting Astronomer, Kitt Peak National Observatory, National Optical Astronomy Observatory, which is operated by the Association of Universities for Research in Astronomy, Inc. (AURA) under cooperative agreement with the National Science Foundation. 
\label{KPNO}}
\addtocounter{address}{1}
\altaffiltext{\theaddress}{Princeton University Observatory, Princeton,
NJ 08544
\label{Princeton}}
\addtocounter{address}{1}
\altaffiltext{\theaddress}{
Department of Physics and Astronomy, The Johns Hopkins University,
Baltimore, MD 21218
\label{JHU}}
\addtocounter{address}{1}
\altaffiltext{\theaddress}{Department of Astronomy, University of California, Berkeley, CA, 94720; Hubble Fellow
\label{Berkeley}}
\addtocounter{address}{1}
\altaffiltext{\theaddress}{Physics Department, University of California, Davis,
CA 95616
\label{UCDavis}}
\addtocounter{address}{1}
\altaffiltext{\theaddress}{IGPP/Lawrence Livermore National Laboratory, Livermore,
CA 94550
\label{IGPP}}
\addtocounter{address}{1}
\altaffiltext{\theaddress}{Space Telescope Science Institute, Baltimore, MD 21218
\label{STScI}}
\addtocounter{address}{1}
\altaffiltext{\theaddress}{Department of Astronomy and Astrophysics,
The Pennsylvania State University,
University Park, PA 16802
\label{PSU}}
\addtocounter{address}{1}
\altaffiltext{\theaddress}{Apache Point Observatory, P. O. Box 59,
Sunspot, NM 88349-0059
\label{APO}}
\addtocounter{address}{1}
\altaffiltext{\theaddress}{Institute for Cosmic Ray Research, University of
Tokyo, Midori, Tanashi, Tokyo 188-8502, Japan
\label{CosmicRay}}
\addtocounter{address}{1}
\altaffiltext{\theaddress}{Mt. Suhora Observatory, Cracow Pedagogical University, ul. Podchorazych 2,
30-084 Cracow, Poland
\label{Cracow}}
\addtocounter{address}{1}
\altaffiltext{\theaddress}{Fermi National Accelerator Laboratory, P. O. Box 500, Batavia, IL 60510
\label{FNAL}}
\addtocounter{address}{1}
\altaffiltext{\theaddress}{Max-Planck-Institut f\"{u}r extraterrestrische Physik, Postfach 1603, 85750 Garching, Germany
\label{MPE}}

\begin{abstract}
We present the discovery of seven quasars
at $z>5.7$, selected from $\sim 2000$ deg$^2$ of 
multicolor imaging data of the Sloan Digital Sky Survey (SDSS).
The new quasars have redshifts $z$ from 5.79  to 6.13.
Five are selected as part of a complete flux-limited sample
in the SDSS Northern Galactic Cap; two have larger photometric errors  and are not part
of the complete sample. One of the new quasars, SDSS J1335+3533 ($z=5.93$),
exhibits no emission lines;
the 3-$\sigma$ limit on the rest-frame equivalent width of
Ly$\alpha$+NV line is 5 \AA.
It is the highest redshift lineless quasar known,
and could be a gravitational lensed galaxy, a BL Lac object or a new
type of quasar. Two new $z>6$ quasars, SDSS 1250+3130 ($z=6.13$) and
SDSS J1137+3549 ($z=6.01$), show deep Gunn-Peterson absorption gaps in 
Ly$\alpha$. These gaps are  narrower the complete Gunn-Peterson absorption troughs observed among quasars at $z>6.2$
and do not have complete Ly$\beta$ absorption.
\end{abstract}

\keywords{quasars: general; quasars: emission line; quasars: absorption lines}
\section{Introduction}

This paper is the fourth in a series presenting $i$-dropout
$z\gtrsim 5.7$ quasars selected from the multicolor imaging data
of the Sloan Digital Sky Survey (SDSS; \cite{York00}, Stoughton et al. 2002). 
In \cite{z58} and in the first three papers of this series
(\cite{PaperI}, Paper I; \cite{PaperII}, Paper II; \cite{PaperIII}, Paper III),
we presented the discovery of twelve luminous quasars
at $z=5.74 - 6.42$, selected from $\sim 4600$ deg$^2$ of 
SDSS imaging in the Northern Galactic Cap.
In this paper, we describe the discovery of 
seven new quasars at $z=5.79 - 6.13$, 
selected from $\sim 2000$ deg$^2$ of new SDSS imaging data.
The scientific objectives, photometric data reduction, candidate
selection and additional photometric and spectroscopic observation
procedures are described
in detail in Paper I and will not be repeated here.
We present the photometric observations of the $i$-dropout
candidates in the new area in \S 2. 
The spectroscopic observations 
and the photometric and spectroscopic properties
of the newly-discovered quasars are described in \S 3. 
SDSS J133550.81+353315.8\footnote{The IAU naming convention for SDSS
sources is SDSS JHHMMSS.SS$\pm$DDMMSS.S, and the positions are expressed in
J2000.0 coordinates. The astrometry is accurate to better than $0.1''$
in each coordinate.} 
 ($z=5.93$, SDSS J1335+3533 for brevity) is a quasar without detectable emission lines; 
we discuss the properties of this unusual object in \S4.

Following the previous papers in this series, we use two cosmologies to
present our results:
(1) $\rm H_0 = 50\ km\ s^{-1}\ Mpc^{-1}$, $\Omega_{\Lambda} = 0$ and
$\Omega_{M} = 1$ ($\Omega$-model); (2)
$\rm H_0 = 71\ km\ s^{-1}\ Mpc^{-1}$, $\Omega_{\Lambda} = 0.73$ and
$\Omega_{M} = 0.27$ ($\Lambda$-model, \cite{Spergel03}).

\section{Candidate Selection and Identification}
The Sloan Digital Sky Survey is using
a dedicated 2.5m telescope (\cite{Gunn05}) and a large format CCD camera (\cite{Gunnetal})
at the Apache Point Observatory in New Mexico
to obtain images in five broad bands ($u$, $g$, $r$, $i$ and $z$,
centered at 3551, 4686, 6166, 7480 and 8932 \AA, respectively; \cite{F96},
Stoughton et al. 2002)
of high Galactic latitude sky in the Northern Galactic Cap.
About $7400$ deg$^2$ of sky have been imaged at the
time of this writing (August 2005).
The imaging data are processed 
with a series of pipelines (Lupton et al. 2001, Pier et al. 2003), resulting in
astrometric calibration errors of $< 0.1''$ rms per coordinate, and
photometric calibration to better than 0.03 mag
(Hogg et al. 2001, Smith et al. 2002, Ivezi{\' c} et al. 2004, Tucker et al. 2005).
These data have been made publicly
available in a series of data releases (EDR: Stoughton et al. 2002; DR1 -- DR4:
Abazajian et al. 2003, 2004, 2005, Adelman-McCarthy et al. 2005).

Quasars at redshifts larger than 5.7  have $i-z>2.2$ in the SDSS filter system, and
become $i$-dropout objects with weak or no detection in all but
the reddest ($z$) band.
In Papers I - III, we presented results from a survey of $i$-dropout
quasar candidates selected from $\sim$ 4600 deg$^2$ of high Galactic
latitude sky in the SDSS Main Survey area, observed in the Springs of
2000 - 2003. 
In the Springs of 2004 and 2005, we continued the search
in 92 new SDSS imaging runs.
These imaging data were taken between May 2003 (Run 3926) and April 2005 (Run 5237),
and cover $\sim 2400$ deg$^2$ of the sky. 
We applied the same color selection criteria as in Paper II (see Figures 1 and
2 in Paper II) to the new SDSS imaging data to select $z>5.7$ quasar
candidates.
A total of $\sim 230$ $i$-dropout candidates that
satisfy the color selection criteria 
(with $z$-band photometric error $\sigma(z) < 0.10$) were selected in the main survey area.
These candidates were selected as part of a flux-limited complete sample (Paper III).
We have also selected a number of fainter candidates with larger $z$-band photometric
error $\sigma(z) = 0.10 - 0.12$,
although they are not part of the complete sample.

The photometric and spectroscopic observations
were carried out over a number of nights between January 2004 and June 2005.
A total of 198 candidates (85\% of those selected) from the complete sample
were observed; thus
the effective area of this sample is $\sim 2000$ deg$^2$.
In addition, data on about thirty fainter candidates with $\sigma(z) = 0.10 - 0.12$
were also acquired.
We first obtained independent $z$ photometry to eliminate false
detections due to cosmic rays and to improve the $i-z$ color
measurement, using the
Seaver Prototype Imaging Camera (SPICAM) on
the ARC 3.5m telescope at the Apache Point Observatory.
Unfortunately, poor
weather limited the number of objects we were able to observe in the
$z$ band. 

Additional $J$-band photometry allows separation of  $z\sim 6$ quasars
from L/T dwarfs, which have similar $i-z$ colors but much redder $z-J$ colors (e.g. Paper I).
$J$-band observations  were carried out using  a number of IR imagers:
(1) the $256 \times 256$ NICMOS imager on Steward Observatory's 2.3m Bok Telescope at Kitt Peak; 
(2) GRIM II (the near infrared GRIsm spectrometer and IMager), on the
ARC 3.5m, which we used through December 2004; 
(3) NIC-FPS (the Near-Infrared Camera/Fabry-Perot Spectrometer), 
also on the ARC 3.5m, which replaced GRIM II in December 2004;
and (4) PANIC (Persson's Auxiliary Nasmyth Infrared Camera, \cite{PANIC})
on the 6.5m Baade Telescope at Las Campanas Observatory.

Spectroscopic observations were carried out over a number of observing runs
between January 2004 and June 2005
using: (1)  the Red-Channel Spectrograph on the MMT 6.5-m telescope on Mt. Hopkins,
(2) the Multi-Aperture Red Spectrograph (MARS) on the 4-m telescope on Kitt Peak,
and (3) the Echellette Imaging Spectrograph (ESI, \cite{ESI}) on the
Keck II telescope on Mauna Kea, Hawaii. 

\section{Discovery of Seven New Quasars at $z>5.7$}

Among the 198 $i$-dropout candidates with $\sigma(z) < 0.10$
that we have observed,
52 are false $z$-band only detections which are most likely cosmic rays;
129 are M or L dwarfs (mostly classified
photometrically based on their red $z - J$ colors);
and 12 are likely T dwarfs (Chiu et al. 2005).
Several objects still lack
proper infrared spectroscopy, so the T dwarf classification
is still preliminary. 
We have not yet obtained $J$-band photometry for the remaining candidates.
Therefore the quasars reported in this paper do not form a complete sample.
Among the faint ($\sigma(z) >0.10$) candidates that are not identified as quasars, 
roughly half are likely cosmic
rays and half are late-type stars, although the larger photometric error does not
allow accurate typing.

Five of the $i$-dropout candidates with
$\sigma(z) < 0.10$ and
two faint candidates with $0.10 < \sigma(z) < 0.12$ are identified as quasars at $z=5.79 - 6.13$.
The finding charts  of the seven new quasars
are presented in Figure 1.
Their spectra are flux-calibrated to match the observed $z$-band
photometry and are shown in Figure 2.
Table 1 presents the photometric properties of the new quasars,
and Table 2 presents the measurements of their continuum properties.
Following Papers I-III, the quantity
$AB_{1280}$ is defined as the AB magnitude of the continuum
at rest-frame 1280 \AA, after correcting for interstellar
extinction using the map of \cite{Schlegel98}.
We extrapolate the continuum to rest-frame 1450 \AA,
assuming a continuum shape $f_\nu \propto \nu^{-0.5}$, to
calculate $AB_{1450}$.
None of the seven quasars is detected in the FIRST (\cite{FIRST}) 
or NVSS (\cite{NVSS}) radio surveys.
FIRST survey has typical 5-$\sigma$ flux limit of $\sim 1$mJy at 20cm.
Assuming a power law continuum of $f_{\nu} \propto \nu^{-0.5}$ in
both radio and UV wavelengths,
we find that quasars in this paper have radio loudness parameter
$R = F_{\rm 5 GHz} / F_{4400}$ smaller than 10, the usual division between
radio-loud and radio-quiet objects (e.g. White et al. 2000).
We have checked the publicly available archival X-ray coverage of these seven
quasars, focusing on observations made by sensitive X-ray imaging detectors.
SDSS~J0927+2001 and SDSS~J1250+3130 lie within pointed X-ray observations
made by the {\it Einstein\/} IPC (1.7~ks) and {\it ROSAT\/} PSPC (0.7~ks),
respectively, but no X-ray detections were obtained. The relatively short
exposures and significant off-axis angles of these two quasars do not
allow physically tight constraints to be placed upon their X-ray-to-optical
flux ratios (see Strateva et~al. 2005). The other five quasars only have
imaging X-ray coverage in the {\it ROSAT\/} All-Sky Survey, and no detections
were obtained, indicating 3-$\sigma$ upper limits 
of X-ray flux of  $\sim \rm 3.1 - 5.7 \times 10^{-13}\
erg\ cm^{-2}s^{-1}$ in the 0.1 -- 2.4 keV band.
Results from a recent {\it Chandra\/} observation of
SDSS~J0840+5624 will be reported in O. Shemmer et~al., in preparation.
The discovery of these seven new quasars brings the total number of $z > 5.7$
quasars from the SDSS to nineteen.

\subsection{Notes on Individual Objects}

\noindent
{\bf SDSS J081827.40+172251.8 ($z=6.00\pm 0.02$).}
This quasar was discovered using the Keck II telescope on April 12, 2005.
Figure 2 shows a spectrum with 20 minute exposure time using Keck/ESI,
smoothed to a resolution of 1800.
SDSS J0818+1722 has very weak emission lines: the rest-frame
equivalent width of Ly$\alpha$+NV is $\sim 10$ \AA, compared
to an average of $\sim 75$ \AA\ for high-redshift quasars (Schneider et al. 1991,
Fan et al. 2001a, Paper III).
The current spectrum has moderate S/N; Ly$\alpha$+NV is the only
detected line.  The redshift is determined from the fit to the
Ly$\alpha$+NV complex.  The uncertainty in this redshift comes from
two sources.  The first is due to absorption in the blue end of the
line due to absorption from the Ly$\alpha$ forest.  The second is the
offset in the redshift determined from Ly$\alpha$ and from high
ionization lines such as CIV on one hand, and that from low ionization
lines such as MgII on the other.  Richards et al. (2002b) use a large
sample of SDSS quasars to show that CIV is blueshifted from MgII by
$824\pm 511$ km s$^{-1}$, corresponding to an uncertainty of $\sim
0.02$ at $z\sim 6$.  Following Papers I-III, we adopt an error bar of
0.02 in the quasar redshift to reflect these uncertainties.

There is also a strong absorption line at $\sim 8870$ \AA\ which is resolved
into a doublet in the full resolution spectrum.
We tentatively identify this feature as the CIV$\lambda\lambda$1548,1551 doublet
at a redshift of 4.726.

\noindent
{\bf SDSS J084035.09+562419.9 ($z=5.85\pm0.02$).}
This object was discovered on January 13, 2004 using the MMT.
Figure 2 shows a  spectrum with 60 minute exposure time using the MMT Red Channel Spectrograph.
The signal-to-noise ratio (S/N) of the spectrum is relatively low. The redshift is estimated
by the best-fit  Ly$\alpha$+NV emission.

\noindent
{\bf SDSS J092721.82+200123.7 ($z=5.79\pm 0.02$).}
This quasar was discovered on May 6, 2005 using the Kitt Peak 4-m telescope.
Figure 2 shows a spectrum with 220 minute exposure time with the MARS
spectrograph on the Kitt Peak 4-meter.
The redshift is estimated by the best-fit  Ly$\alpha$+NV emission.

\noindent
{\bf SDSS J113717.73+354956.9 ($z=6.01\pm0.02$).}
This quasar was discovered on April 27, 2004 using the MMT.
Figure 2 shows a 40 minute Keck/ESI spectrum taken on January 6, 2005.
The spectrum shows clear detections of Ly$\alpha$+NV, OI and SiIV emission.
We adopt the average of the best-fit values of these lines as the quasar redshift.
A strong and marginally resolved absorption feature
is detected at 9180 \AA; we are not able to unambiguously identify this feature.

The absorption spectrum shows a dark absorption gap
in Ly$\alpha$ transition 
immediately blueward of the emission line, where no flux is detected.
Following Becker et al. (2001), Fan et al. (2002) and White et al.
(2003), we define the transmitted flux ratio as:
\begin{equation} 
{\cal T}(z_{abs}) \equiv \left\langle f_\nu^{obs}/f_\nu^{con} \right\rangle,
\end{equation}
where $f_\nu^{con}$ is the continuum level extrapolated from the
red side of the Ly$\alpha$ emission line.
We find an average transmitted flux ${\cal T} = -0.005 \pm 0.004$ for 
 $z_{abs}$ between 5.83 and 5.90. 
The effective Ly$\alpha$ optical depth in this dark gap is $\tau_{\rm eff}>5.5$,
comparable to the limits measured in the complete Gunn-Peterson (1965) troughs
detected in quasars at $z>6.1$ (Becker et al. 2001, Fan et al. 2002, White et al. 2003, Paper III, see also Songaila 2004).
The redshift extent of this dark gap is considerably narrower than those seen in
quasars at $z>6.1$.
Furthermore, flux is clearly detected in the Ly$\beta$ transition
over the same redshift range, with an average transmitted flux
of ${\cal T}_{\beta} = 0.013 \pm 0.003$, indicating that IGM is still highly ionized
in this region, although the mean optical depth has increased substantially
from lower redshift.
In a subsequent paper (X. Fan,   et al. in preparation), we present a full analysis of
the Gunn-Peterson absorption in the spectra of this and the other 18
$z > 5.7$ SDSS quasars. 

\noindent
{\bf SDSS J125051.93+313021.9 ($z=6.13\pm0.02$).}
This quasar was discovered on April 24, 2004 using the MMT.
We did not carry out $J$-band photometry; 
the red $i-z$ color and lack of detection ($J>16.5$) in 2MASS (\cite{2MASS})
suggested that it was likely to be a high-redshift quasar.
Figure 2 shows a 60 minute Keck/ESI spectrum taken on January 6, 2005.
The spectrum shows clear detections of Ly$\alpha$+NV, Ly$\beta$+OVI, 
OI and SiIV emission lines.
We adopt the average of the best-fit values of these lines as the quasar redshift.
It is the highest redshift object presented in this
paper, and the fifth most distant quasar yet discovered.
On the blue wing of the OI emission line, a MgII doublet at $z_{abs} = 2.29$
is clearly detected, with a total rest-frame equivalent width of 4.5 \AA.

In the absorption spectrum of this quasar, we detect a deep Gunn-Peterson
absorption gap at $5.69 < z_{abs} < 5.95$ in Ly$\alpha$ transition.
The average transmitted flux, ${\cal T}= 0.005 \pm 0.003$, indicates
a lower limit of the Gunn-Peterson optical depth
$\tau_{\rm eff} > 5.3$. Similar to the case of
J$1137+3549$, residual flux is clearly detected over the
same redshift range in the Ly$\beta$ transition, with a transmitted flux
of ${\cal T}_{\beta} = 0.008 \pm 0.002$. 

\noindent
{\bf SDSS J133550.81+353315.8 ($z=5.93\pm0.04$).}
This quasar was discovered on June 13, 2005 using the MMT.
Figure 2 shows a 120 minute MMT/Red Channel spectrum.
This object has a $z$-band photometric error $\sigma(z) = 0.11$,
therefore it is not part of the complete flux-limited sample of
$z\sim 6$ quasars.
SDSS J1335+3533 is the most unusual object presented in this paper:
it does not have a detectable Ly$\alpha$+NV emission line.
We discuss its properties and possible interpretations in \S4.

\noindent
{\bf SDSS J143611.74+500706.9 ($z=5.83\pm0.02$).}
This object was discovered on April 28, 2004, using the Kitt Peak 4-m Telescope.
Figure 2 shows a spectrum with 60 minute exposure time using the MMT/Red Channel Spectrograph.
This object has a $z$-band photometric error $\sigma(z) = 0.12$,
and therefore will not be part of the complete flux-limited sample of
$z\sim 6$ quasars.
The redshift is estimated by the best-fit Ly$\alpha$+NV line.

\section{SDSS J1335+3533, a Lineless Quasar at $z=5.93$}

Figure 3 shows the MMT spectrum of SDSS J1335+3533.
This object is clearly at high redshift:
it exhibits a sharp Ly$\alpha$ break at $\lambda < 8410$ \AA,
and a sharp Ly$\beta$ break at $\lambda < 7130$ \AA, indicating
a redshift $z>5.92$. 
However, no strong Ly$\alpha$ emission line is detected.
Although there is a hint of excess flux at the edge of the Ly$\alpha$
break, the spectrum is consistent with a pure power-law continuum. 
Fitting a power-law plus emission line model to 
the spectral region redward of the Ly$\alpha$ break, we find
the 3-$\sigma$ limit on the rest-frame equivalent width of
Ly$\alpha$+NV line to be 5 \AA. 
Most high-redshift quasars are characterized by the presence of strong
Ly$\alpha$ and NV emission lines, and the rest-frame equivalent width of Ly$\alpha$+NV
is $70\pm15$\AA\ for $z\sim 4$ quasars (e.g. Schneider, Schmidt \& Gunn 1991,
Vanden Berk et al. 2001, Fan et al. 2001a, 2004).
Figure 3 also displays the composite spectrum of 16 non-BAL quasars
at $z>5.7$ (excluding SDSS J1335+3533), constructed following
the procedure described in Paper III. The lack of Ly$\alpha$
is striking. The current observations do not have sufficient S/N to place
strong constraints on the presence of OI and SiIV lines; 
another strong emission line, CIV, is now beyond 1$\mu m$ and
will require new infrared spectroscopy.
We base our estimate of the redshift of SDSS J1335+3533 on the size of the proximity
effect zone in the Ly$\alpha$ forest close to the Ly$\alpha$ emission line.
Details of calculating the size of proximity zone around luminous quasars at
$z\sim 6$ are described in a subsequent paper (X. Fan et al. in preparation).
Briefly, in regions of the IGM close to the quasar, transmitted flux 
is enhanced due to increased ionizing flux
from the quasar itself. 
The transmitted flux declines to that in
the average IGM at large discance; at $z>5.7$, ${\cal T} < 0.05$. 
Fluxes in the 16 other $z > 5.7$ non-BAL quasars from the SDSS drop to 20\%
of the continuum level at a redshift $\Delta z = 0.11 \pm 0.04$
blueward of the Ly$\alpha$ emission line.  This
value is a function of luminosity and redshift. 
SDSS J1335+3533 is close to the median redshift and luminosity of our
sample. Including these effects does not change
our redshift estimate.
For this quasar, the
transmitted flux falls to 20\% of the extrapolation of the
long-wavelength flux at $z_abs = 5.82$, implying that the true
redshift of the quasar is $z = 5.93 \pm 0.04$. 
We obtain a similar redshift using the onset of the Ly$\beta$ forest in
the spectrum.  Note that this redshift determination depends somewhat
on the assumed size of the HII region around SDSS J1335+3533.

Lineless quasars are very rare at high redshift.
This is unlikely to be a selection effect.
The SDSS color selection of high-redshift quasars is mostly based
on the presence of a strong Lyman break. It  does not strongly select against quasars with
small Ly$\alpha$ emission line equivalent widths (Fan et al. 1999, Richards et al. 2002a).
The first lineless quasar found at high redshift was SDSS J1532--0030
($z=4.62$, Fan et al. 1999); its spectrum is very similar to that
of SDSS J1335+3533.
Since then, the SDSS has so far discovered about a dozen quasars at $3<z<5$ with
rest-frame Ly$\alpha$ equivalent width smaller than 5 \AA\ (e.g., Anderson et
al. 2001; Collinge et al. 2005).
They appear to form a distinct class of high-redshift {\em lineless} quasars.
SDSS J1335+3533 represents the most distant member of this class of objects.

The nature of these lineless quasars remains a mystery.
They could be  ultra-luminous or lensed 
galaxies, high-redshift BL Lac objects, or a new class of quasar with very weak or absent
broad emission lines.
The absolute AB magnitude of SDSS J1335+3533 at 1450 \AA\ (rest-frame) is $\rm
M_{1450}=-26.81$ ($\Lambda$-model).
Thus, although the S/N of our optical spectrum is probably not sufficient
 to rule out the presence of stellar absorption lines, this
object is much too luminous to be an ordinary galaxy unless it is
strongly lensed.
There is no
indication that SDSS J1335+3533 is extended beyond the point spread function of
the $1.3''$ FWHM SDSS image; this is not yet a strong constraint on the
lensing hypothesis.

BL Lac objects, by definition, show no or very weak emission lines.
They are also characterized by strong radio and X-ray emission,
optical variability, and strong and variable optical polarization
due to synchrotron radiation from the relativistic jet (e.g. \cite{UP95}).
The highest redshift BL Lac in the literature is at $z<2$
(\cite{LM99}).
SDSS J1335+3533 is not detected in the FIRST radio survey;
the $4\sigma$
upper limit of 0.8 mJy/beam at this position, including CLEAN bias (White et al.
1997). This  implies a radio-to-optical spectral index in the observed frame of
$\alpha_{ro} < 0.26$ (4-$\sigma$).
Classical radio-selected BL Lacs have $\alpha_{ro} > 0.3$ (Stocke et al. 1990),
while X-ray selected BL Lacs can have $\alpha_{ro}$ as small as $\sim 0.1$
(\cite{LM99}).
Collinge et al. (2005) present a sample of 386 BL Lac candidates identified
from the SDSS by their lack of emission lines. Most of them are at
low redshift ($z < 1$), and only $\sim 3\%$ of these optically-selected candidates 
have $\alpha_{ro} < 0.2$. 
SDSS J1335+3533 is also not detected on the ROSAT full-sky pixel images 
(\cite{ROSAT}),
implying a 3$\,\sigma$ upper limit of X-ray flux of  $\sim \rm 3.7 \times 10^{-13}\
erg\ cm^{-2}s^{-1}$ in the 0.1 -- 2.4 keV band, and an optical-to-X-ray
index in the observed frame $\alpha_{ox} > 1.1$ at
3-$\sigma$. This limit does not yet place strong constraints on the
BL Lac hypothesis, as it is in the middle of  $\alpha_{ox}$ distribution
for X-ray selected BL Lacs (\cite{Stocke90}, Sambruna et al. 1996).

The current observations cannot yet place strong constraints on the
nature of SDSS J1335+3533: whether it is a lensed galaxy, a very high redshift
BL Lac, or a new type of quasar with no or a very weak emission line region.
Further observations, including high-resolution imaging to test
the lensing hypothesis, optical monitoring for variability, deep radio and
X-ray observations
(e.g. Schneider et al. 2003), and optical polarimetric measurements to test
the BL Lac hypothesis,  are needed. 
Spitzer observations of such objects in the mid-infrared will probe the presence of hot
dust in the vicinity of the object, differentiating the hypotheses that the
source is AGN-powered, star-formation powered, or beamed (A. M. Diamond-Stanic, et al.
in preparation).
Stalin \& Srianand (2005) suggest that such objects could be
explained by microlensing of the continuum source that only boosts
the continuum; if this were the case, long-term monitoring would show a 
decreasing continuum and a detectable Ly$\alpha$ emission line 
after the lensing event passes.

\section{Future Observations}

Combining the seven new quasars presented in this paper with those
reported in Papers I-III, we have a sample of 19 quasars at $z>5.7$.
Among them, fourteen are selected using a uniform set of color selection
criteria from an effective area of $\sim 6600$ deg$^2$, forming
a flux-limited sample.
Because the spectroscopic observations of $i$-dropout candidates
have not been completed,
we do not derive statistical properties
of the sample in this paper.
In Spring 2005, the SDSS
essentially completed its imaging survey in the Northern Galactic
Cap, with a total area of $\sim 7550$ deg$^2$
In Spring 2006, we will continue to search for $i$-dropout quasar candidates
in the remaining new area, as well as observe additional candidates
in the previously studied area that meet the selection criteria because of improved photometric calibrations.
Combining current data with these planned observations, we will be able to define a sample of $z>5.7$
quasars covering the entire SDSS Northern Galactic Cap area, and use it to 
refine the measurements of  the evolution of the luminosity function and emission line properties
of the highest redshift quasars.

    Funding for the creation and distribution of the SDSS Archive has been provided by the Alfred P. Sloan Foundation, the Participating Institutions, the National Aeronautics and Space Administration, the National Science Foundation, the U.S. Department of Energy, the Japanese Monbukagakusho, and the Max Planck Society. The SDSS Web site is http://www.sdss.org/.
    The SDSS is managed by the Astrophysical Research Consortium (ARC) for the Participating Institutions. The Participating Institutions are The University of Chicago, Fermilab, the Institute for Advanced Study, the Japan Participation Group, The Johns Hopkins University, the Korean Scientist Group, Los Alamos National Laboratory, the Max-Planck-Institute for Astronomy (MPIA), the Max-Planck-Institute for Astrophysics (MPA), New Mexico State University, University of Pittsburgh, University of Portsmouth, Princeton University, the United States Naval Observatory, and the University of Washington.
We thank the staffs at Apache Point Observatory, the MMT, the Bok
Telescope, Kitt Peak, Keck and Magellan for their
expert help.
We acknowledge support from NSF grant AST-0307384, a Sloan Research Fellowship
and a Packard Fellowship for Science and Engineering (X.F.), and NSF grants
AST-0071091 and AST-0307409 (M.A.S.).

\newpage

\begin{figure}[hbt]
\plotone{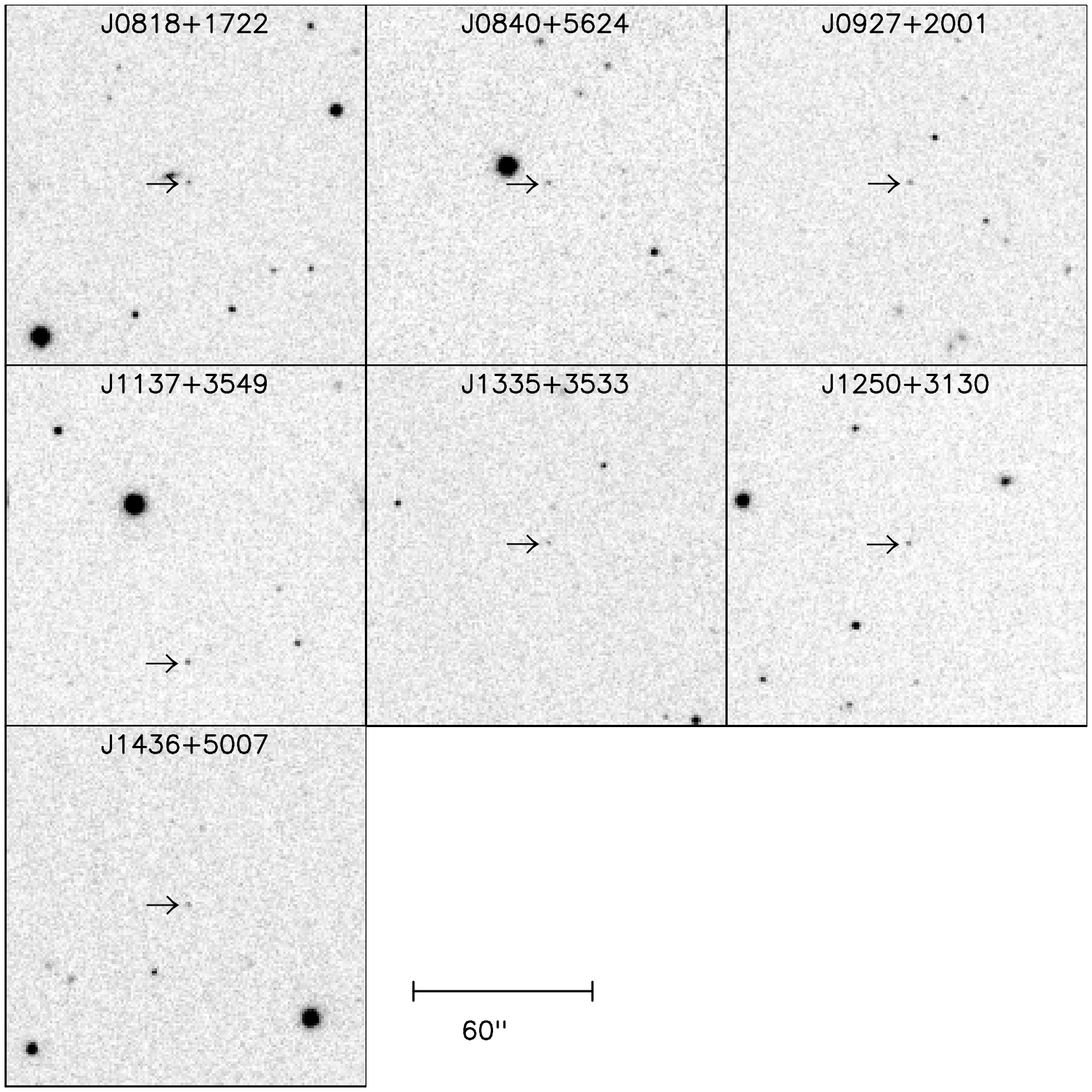}
\caption{SDSS $z$-band images of the seven new $z>5.7$ quasars.
Each side of the finding chart is 120$''$. 
North is up and East is left.
}
\end{figure}

\begin{figure}[hbt]
\plotone{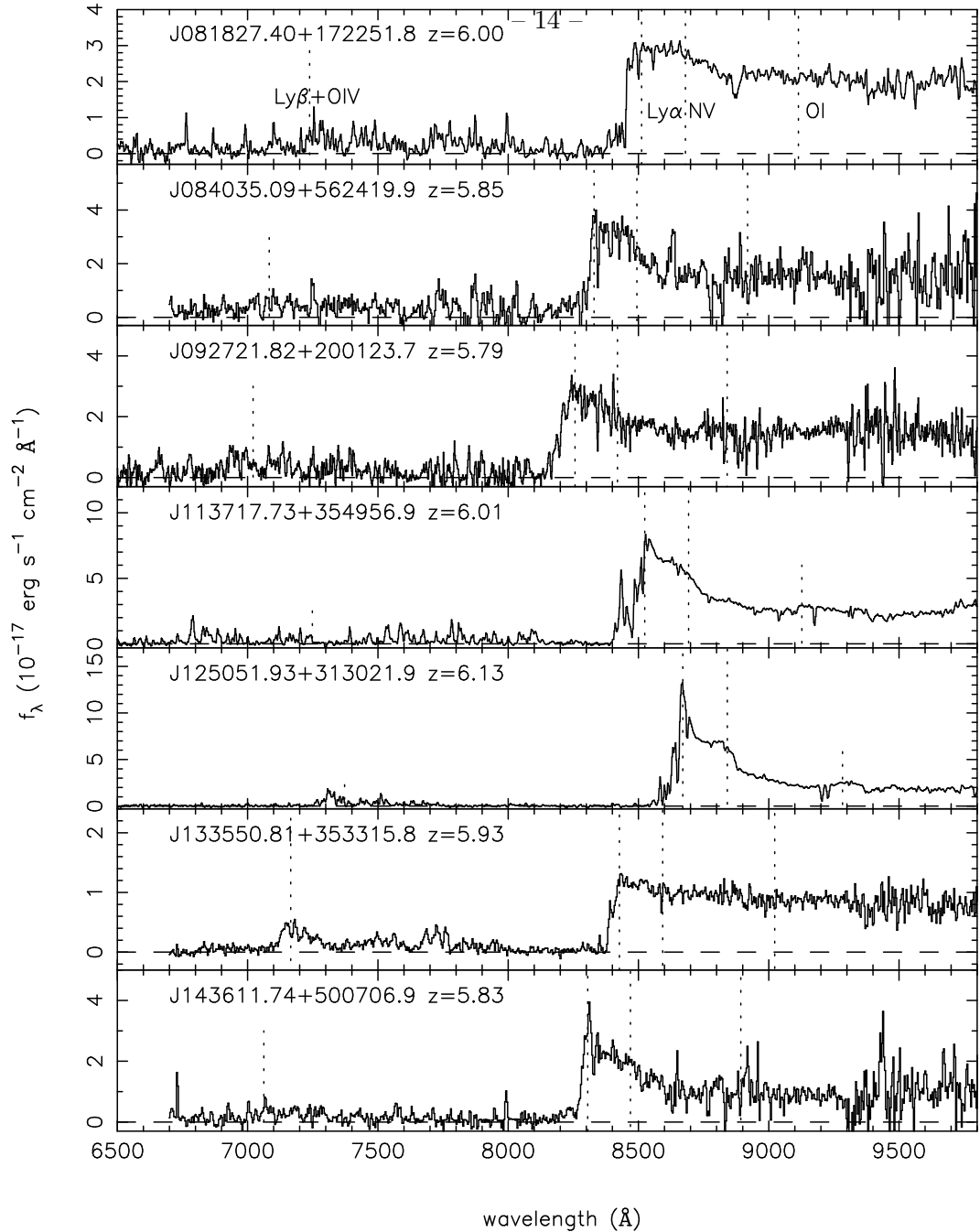}
\caption{Spectra of the seven new quasars at $z>5.7$.
The spectral resolutions
are between 500 and 4000, depending on the spectrograph used.
Vertical lines indicate wavelengths at which major emission lines are expected.
}
\end{figure}

\begin{figure}[hbt]
\plotone{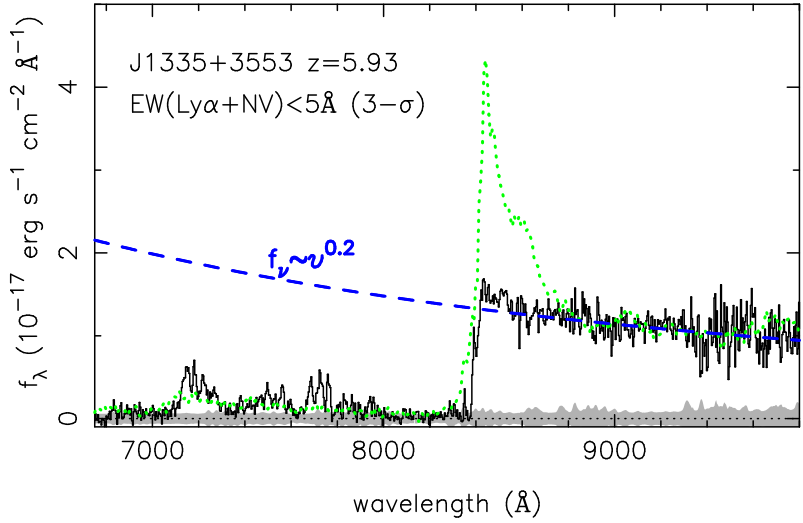}
\caption{
Spectrum of the lineless quasar SDSS J1335+3533 ($z=5.93$).
The shaded area around zero flux shows the 1-$\sigma$ error array.
The dashed line is the best-fit power law. The absolute value of power-law
index is uncertain by $\sim 0.3$ due to possible error in the spectrophotometric
calibration. The dotted line shows the composite spectrum of
16 quasars at $z>5.7$ (arbitrary normalization). The Ly$\alpha$+NV emission line has an average rest-frame
equivalent width of $\sim 70$ \AA\ at $z>4$, while for
SDSS J1335+3533, it has a 3-$\sigma$ upper limit of 5 \AA.
}
\end{figure}

\begin{deluxetable}{cccclc}
\tablenum{1}
\tablecolumns{6}
\tablecaption{Photometric Properties of Seven New $z>5.7$ Quasars}
\tablehead
{
Quasar & redshift & $i$ & $z$ & $J$ & SDSS run\\
(SDSS) & & & & & 
}
\startdata
J081827.40+172251.8  & 6.00 $\pm$ 0.02  & 22.19  $\pm$ 0.22  & 19.60  $\pm$ 0.08   & 18.54 $\pm$ 0.05  & 5045 \\
J084035.09+562419.9  & 5.85 $\pm$ 0.02  & 22.43  $\pm$ 0.34  & 19.76 $\pm$  0.10   & 19.00 $\pm$ 0.10  & 4204  \\
J092721.82+200123.7  & 5.79 $\pm$ 0.02  & 22.12  $\pm$ 0.17  & 19.88  $\pm$ 0.08   & 19.01 $\pm$ 0.10  & 5138  \\
J113717.73+354956.9  & 6.01 $\pm$ 0.02  & 22.57  $\pm$ 0.30  & 19.54  $\pm$ 0.07   & 18.41 $\pm$ 0.05  & 4392  \\
J125051.93+313021.9  & 6.13 $\pm$ 0.02  & 22.14  $\pm$ 0.18  & 19.53  $\pm$ 0.08   & $>16.5$   & 4623 \\
J133550.81+353315.8  & 5.93 $\pm$ 0.04  & 22.67  $\pm$ 0.99  & 20.10  $\pm$ 0.11   & 18.97 $\pm$ 0.05  & 4470 \\ 
J143611.74+500706.9  & 5.83 $\pm$ 0.02  & 22.76  $\pm$ 0.28  & 20.00  $\pm$ 0.12   & 19.04 $\pm$ 0.10 & 3180 
\enddata
\tablenotetext{}{The SDSS photometry ($i,z$) is
reported in terms of {\em asinh magnitudes} on the AB system.
The asinh magnitude system is defined by Lupton, Gunn \& Szalay (1999);
it becomes a linear scale in flux when the absolute value of the
signal-to-noise ratio is less than about 5. In this
system, zero flux corresponds to 24.4 and 22.8,
in $i$, and $z$, respectively.
The $J$ magnitude is on a Vega-based system.}
\end{deluxetable}

\begin{deluxetable}{ccccccc}
\tablenum{2}
\tablecolumns{7}
\tablecaption{Continuum Properties of new $z>6$ Quasars}
\tablehead
{
Quasar & redshift & $AB_{1280}$ & $AB_{1450}$ & $M_{1450}$ & $M_{1450}$ & $E(B-V)$ \\
(SDSS)  &  &   &  & ($\Lambda$-model) & ($\Omega$-model) &  (Galactic)
}
\startdata
J081827.40+172251.8 & 6.00 & 19.41 & 19.34 & $-27.37$ & $-27.14$ & 0.04  \\
J084035.09+562419.9 & 5.85 & 20.10 & 20.04 & $-26.64$ & $-26.40$ & 0.04  \\
J092721.82+200123.7 & 5.79 & 19.94 & 19.87 & $-26.78$ & $-26.55$ & 0.03  \\
J113717.73+354956.9 & 6.01 & 19.70 & 19.63 & $-27.08$ & $-26.85$ & 0.02  \\
J125051.93+313021.9 & 6.13 & 19.71 & 19.64 & $-27.11$ & $-26.87$ & 0.02  \\
J133550.81+353315.8 & 5.93 & 19.96 & 19.89 & $-26.81$ & $-26.58$ & 0.01  \\
J143611.74+500706.9 & 5.83 & 20.23 & 20.16 & $-26.51$ & $-26.28$ & 0.02 
\enddata
\end{deluxetable}

\end{document}